# Astronomical image processing based on fractional calculus: the *AstroFracTool*

**Roberto Marazzato**[1] and **Amelia Carolina Sparavigna**[2]
1 Department of Automation and Computer Science, Politecnico di Torino, Torino, Italy
2 Department of Physics, Politecnico di Torino, C.so Duca degli Abruzzi 24, Torino, Italy

**Abstract**
The implementation of fractional differential calculations can give new possibilities for image processing tools, in particular for those that are devoted to astronomical images analysis. As discussed in arxiv:0910.2381, the fractional differentiation is able to enhance the quality of images, with interesting effects in edge detection and image restoration. Here, we propose the *AstroFracTool*, developed to provide a simple yet powerful enhancement tool-set for astronomical images. This tool works evaluating the fractional gradient of an image map. It can help produce an output image useful for further research and scientific purposes, such as the detection of faint objects and galaxy structures, or, in the case of planetary studies, the enhancement of surface details.

**1. Introduction**
Digital images are arrays of numbers that can be manipulated by computer software. Using for instance the RGB colour model, that is, the additive colour model in which the addition of red, green and blue lights reproduces the colours, we associate to each pixel of the image three numbers ranging form 0 to 255, the colour tones. We can then prepare a code in a programming language to analyse this array of numbers and prepare an output map corresponding to our specific desired evaluations.
There are many image processing resources, most of them freely available and quite friendly to use, which can be useful in manipulating images. In spite of this abundance, the development of new methods and tools is still worthy of efforts. Here, we propose the *AstroFracTool*, developed to provide an enhancement tool-set for astronomical images. This tool works evaluating the fractional gradient of an image map, that is, it works by means of a fractional differentiation. Let us note that, to the authors' knowledge, none among the free digital imaging software packages uses routines based on fractional calculus.
Fractional calculus provides derivation and integration of functions to non-integer order [1-3]. The problem is rather old, as shown by a correspondence between Leibniz and L'Hopital [4]. The fact that we are not familiar with fractional calculus is due to its development in the field of pure mathematics [5]. First applications were proposed in 1920. Only recently, it was approached in image processing [6], where it can be rather interesting for filtering and edge detection [7-9]. As proposed in [7] and discussed in [10], fractional differentiation is suitable for edge detection and for enhancing the image quality. In [11], the fractional differentiation was used for processing astronomical images.
The recording of astronomical images is characterised by very long exposure times, often of many hours, or on the recording a movie. The image is prepared stacking many frames of the sequence. Long time exposure photography suffers from many sources of noise, due to surface lights and flickering of atmosphere. This noise remains recorded in the resulting image. In the case that a stacking procedure is used, the level of noise increases when each image is stacked on [11]. Removing the noise has the consequence to deplete the image of detail and then, in the final image, faint objects are fainter and edges are smoother.

As discussed in [10,11], fractional differentiation can help scan and examine an astronomical image: there, images were processed with a Fortran code running on a Unix machine. Here we propose the package *AstroFracTool* as the toll, suitable to improve the astronomical images, running on Windows.

**2. The algorithm**

*AstroFracTool* is based on the discrete implementation of the fractional gradient as in Ref.[12,13]. Let $\nu$ be a real number. If we have a two-dimensional function $s(x,y)$, where $x,y$ can have only discrete values, that is $x=1,2,...,n_x$ and $y=1,2,...,n_y$, partial derivatives can be evaluated as:

$$\frac{\partial^\nu s(x,y)}{\partial x^\nu} = s(x,y) + (-\nu)s(x-1,y) + \frac{(-\nu)(-\nu+1)}{2}s(x-2,y) +$$
$$+ \frac{(-\nu)(-\nu+1)(-\nu+2)}{6}s(x-3,y) + ...$$
$$\frac{\partial^\nu s(x,y)}{\partial y^\nu} = s(x,y) + (-\nu)s(x,y-1) + \frac{(-\nu)(-\nu+1)}{2}s(x,y-2) +$$
$$+ \frac{(-\nu)(-\nu+1)(-\nu+2)}{6}s(x,y-3) + ...$$
(1)

The fractional gradient is defined as [13]:

$$grad^\nu = \frac{\partial^\nu}{\partial x^\nu}\mathbf{u}_x + \frac{\partial^\nu}{\partial y^\nu}\mathbf{u}_y = G_x^\nu \mathbf{u}_x + G_y^\nu \mathbf{u}_y \qquad (2)$$

$\mathbf{u}_x, \mathbf{u}_y$ are the unit vectors of the two-dimensional space. The magnitude is $G^\nu = ((G_x^\nu)^2 + (G_y^\nu)^2)^{1/2}$. In the case when the fractional order parameter is $\nu=1$, we have the well-know gradient. $\nu=0$ gives the original function.

*AstroFracToll* evaluates the magnitude of gradient $G^\nu(x,y,c)$, when function $s(x,y)$ is the image map $b(x,y,c)$ for each colour tone $c$. Partial derivatives contain only the first four terms in Eq.1. For each colour, we find the maximum value $G_{Max}^\nu(c)$ on the image map. After we define the output map as in the following:

$$b_G(x,y,c) = 255 \left( \frac{G^\nu(x,y,c)}{G_{Max}^\nu(c)} \right)^\alpha \qquad (3)$$

where $\alpha$ is a parameter suitable to adjust the image visibility. The role of $\nu,\alpha$ parameters in the image processing is illustrated in Fig.1: note that it is possible to see more details near the edges and inside craters. In Fig.2, we see another example with a galaxy image; in this case, parameter α was set to a fixed value.

The algorithm, in fact, enhances image edges turning out to be useful in studying images with faint grey- or colour tone variations. Therefore, the tool reveals faint objects in the image, increasing then possibilities to discover small erratic bodies.

Let us remember that fractional differentiation behaves differently from that of integer derivatives and then the results we can obtain by applying the fractional gradient are different from those obtained by means of usual image processing tools, such as GIMP, for instance. These programs in fact have filtering actions based on integer order differentiation. GIMP and other tools are suitable for a further processing of the map obtained from fractional gradient evaluation, to have an enhancement of colours, brightness and contrast.

## 3. *AstroFracTool* features

*AstroFracTool* has been developed to provide a simple enhancement tool-set. As previously told, its first release is based on the fractional gradient concept. The tool is working on any BMP or JPG picture of sky objects. The package runs on Windows NT/2K with a .NET package, which can be downloaded free from the MS site.

### 3.1 GUI
The interface is quite simple: it is possible to open a selected image and set the processing parameters (screenshot #1) and choose the image to be displayed (screenshot #2). It is possible to create an HTML report (the use of which is strongly suggested for recording purposes), as the one which is displayed at http://staff.polito.it/roberto.marazzato/pleiades (screenshot #3), editing the most relevant data of the last report.

### 3.2 Features to be added
Here we are discussing and working with the first trial version of *AstroFracTool*. New features will be added soon. The next one is being theoretically analysed, and will allow the user to control each separate colour channel both in the horizontal and in the vertical direction. Suggestions from astronomers, to improve the software according to the needs of the intended users, will be very useful to prepare the new versions.

### 3.3 Availability
*AstroFracTool c*an be freely downloaded at the following URL:
http://staff.polito.it/roberto.marazzato/AstroFract.zip .

## 4. Examples and discussions
With *AstroFracToll* we are able to detect the faint stars in the image background. An example is shown in Fig.3. A proposed application for the program could a use for detection of erratic bodies such as comets or asteroids by comparing images of the same region of space. In the upper part of Fig.3, (3.a) is showing the original image, (3.b) and (3.c) the maps obtained with the fractional gradient. The lower part of the figure shows the same images, processed with GIMP, with increased brightness and contrast. Note that image (3.d) is the best result that we can obtain with GIMP.

Being dependent on local tone variations, the application of a fractional gradient to an image is able to enhance galactic structures, which are depending on the density matter variation. Two examples are shown in Fig.4 and 5. *AstroFracTool* increases both stars and galactic structure visibility. The output image obtained by the software can be improved by increasing brightness and contrast. This further processing does not add or remove information. When brightness and contrast of the input image, which is of the original image, are changed, we often find that some information is lost. For instance, we try to increase the brightness and improve the contrast to have a better stars visibility, but we have, at the same time, that galactic details are removed. Because of processing several astronomical images, we suggest that fractional differentiation could properly enter those image-processing tools devoted to the detection of faint objects in astronomical images.

We have tested the first trial version of *AstroFracTool*. Future works are needed, improving the algorithm with new possibilities. As previously told, an interesting processing feature could be the separate control of colour channels; another one could be the comparison of images (addition and subtraction of images). As a matter of fact, the use of this tool by astronomers will be very useful to prepare the new versions.

**ACKNOWLEDGEMENT**

Authors thank Paul Milligan of the British Astronomical Association, Isle of Man Astronomical Society (http://www.eyetotheuniverse.com/), for interesting discussions and suggestions.

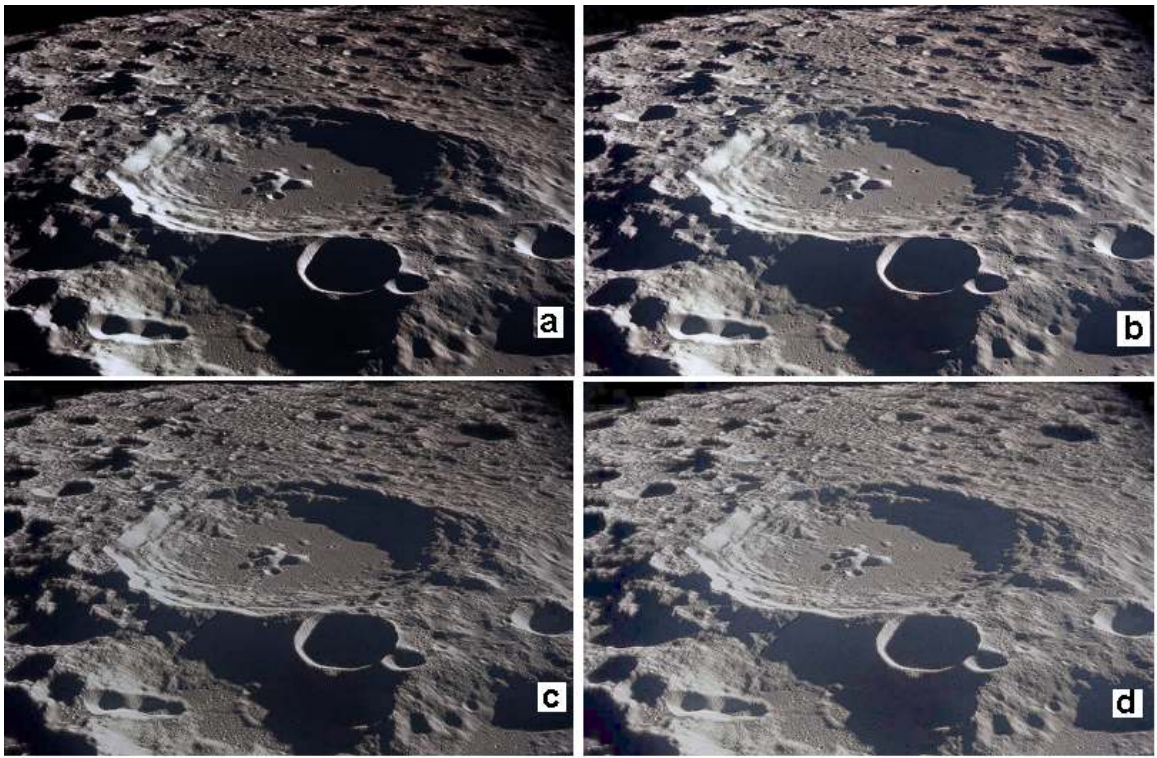

**Figure 1**. The figure shows the role of $\nu, \alpha$ parameters in the image processing. The original image (a), taken by Apollo 11, shows Moon craters. Image (b-d) are the maps obtained by means of *AstroFracTool* with: (b) $\nu = 0.0, \alpha = 0.6$, (c) $\nu = 0.3, \alpha = 0.6$ and (d) $\nu = 0.3, \alpha = 0.4$. The original image is obtained again when $\nu = 0.0, \alpha = 1$.

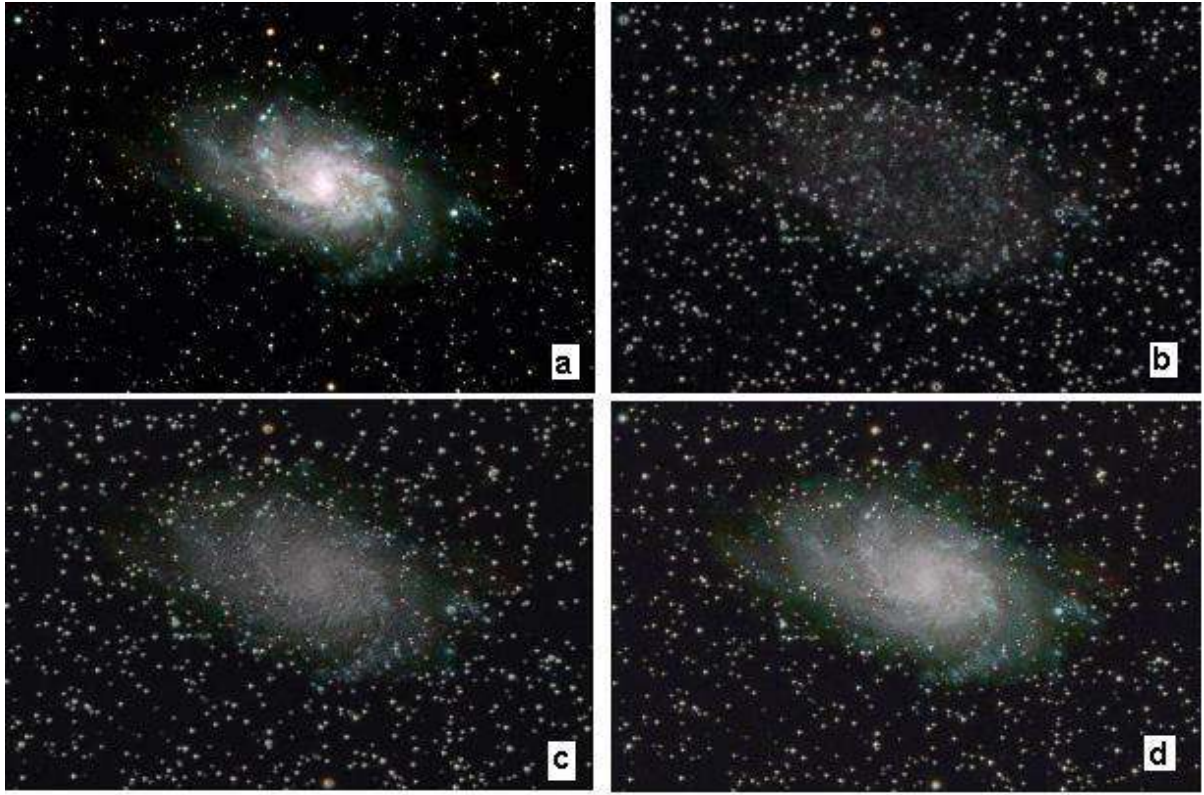

**Figure 2**. Image (a) shows Messier 33 (Triangulum galaxy, author: Paul Milligan, http://www.eyetotheuniverse.com/). (b), (c) and (d) are the maps obtained with $\nu = 1.$, $\nu = 0.7$ and $\nu = 0.4$ respectively. For the three images, we have set $\alpha = 0.5$. See [11] for more details.

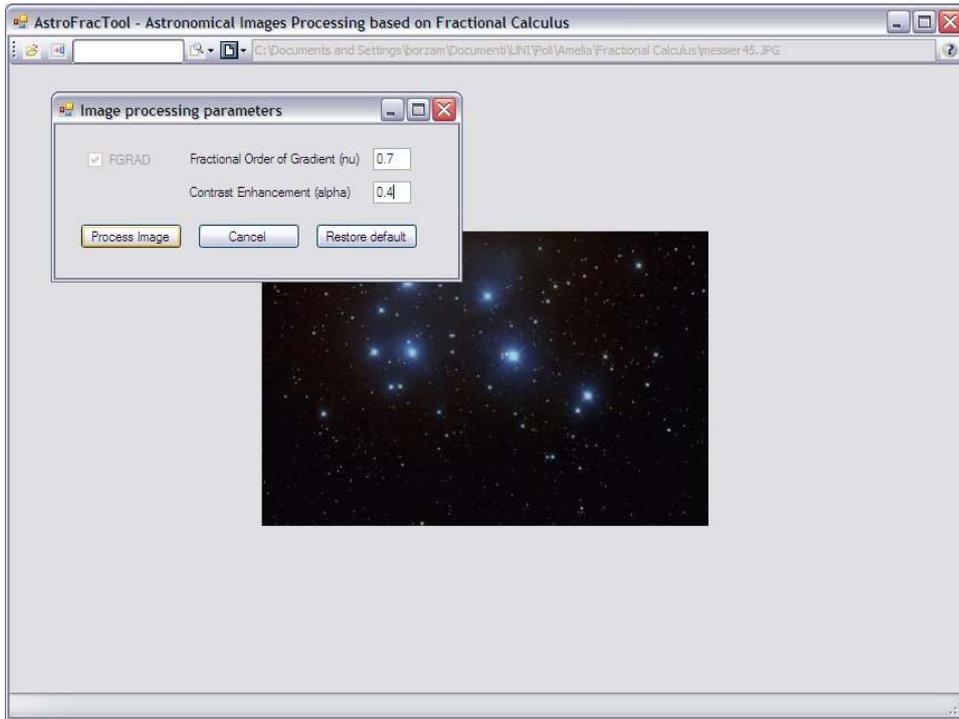
Screenshot #1

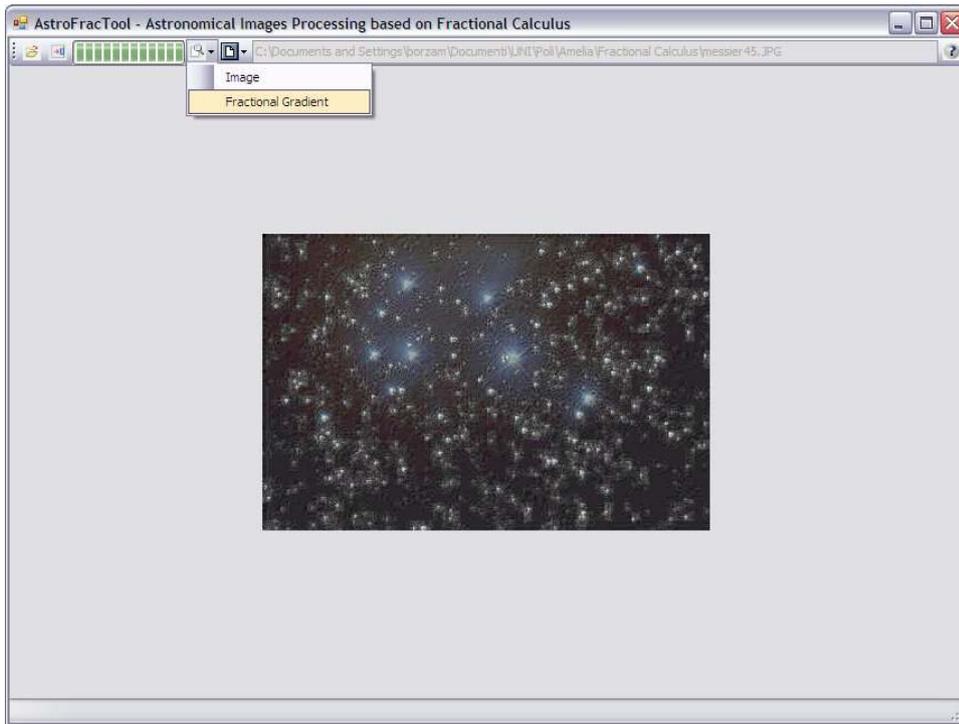
Screenshot #2

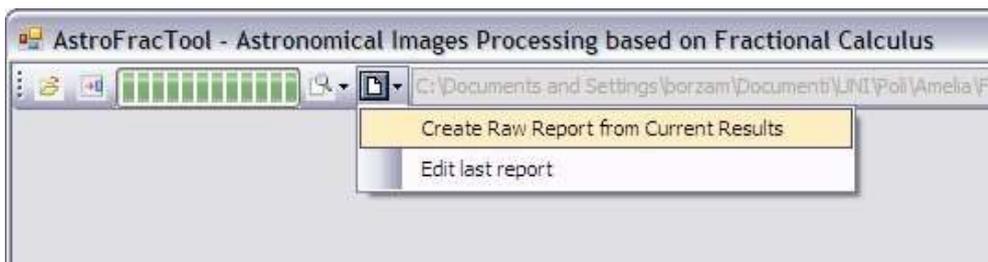
Screenshot #3

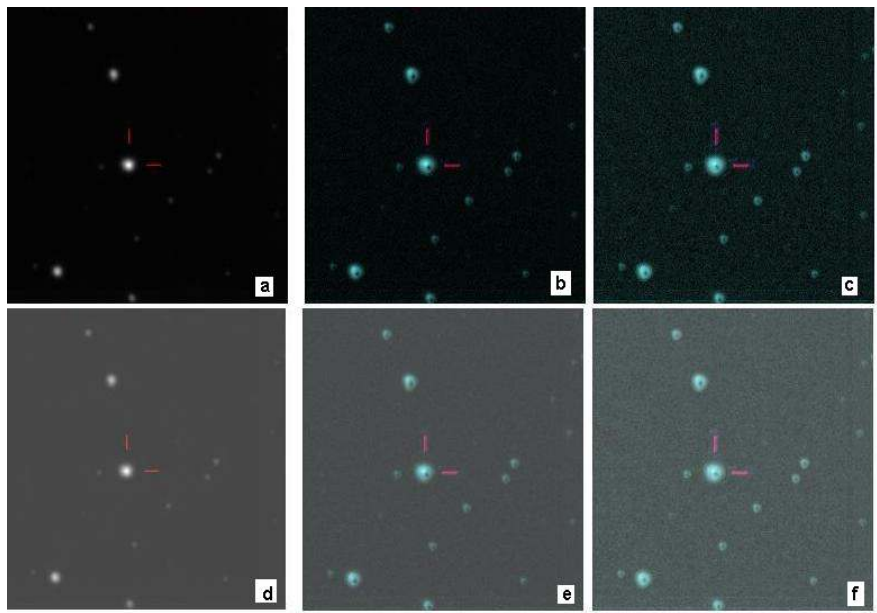

**Figure 3.** In the upper part of Fig.2, (a) is showing the original image, (b) and (c) the maps obtained with the fractional gradient. (b) has $v = 0.7, \alpha = 0.5$ and (c) $v = 0.8, \alpha = 0.5$. The lower part of the figure shows the same images, processed with GIMP, to increase brightness and contrast. Note that image (d) is the best result that we can obtain with GIMP. The original image is published by the Nordic Optical Telescope Scientific Association, authors L.Ø. Andersen, L.Malmgren, F.R. Larsen.

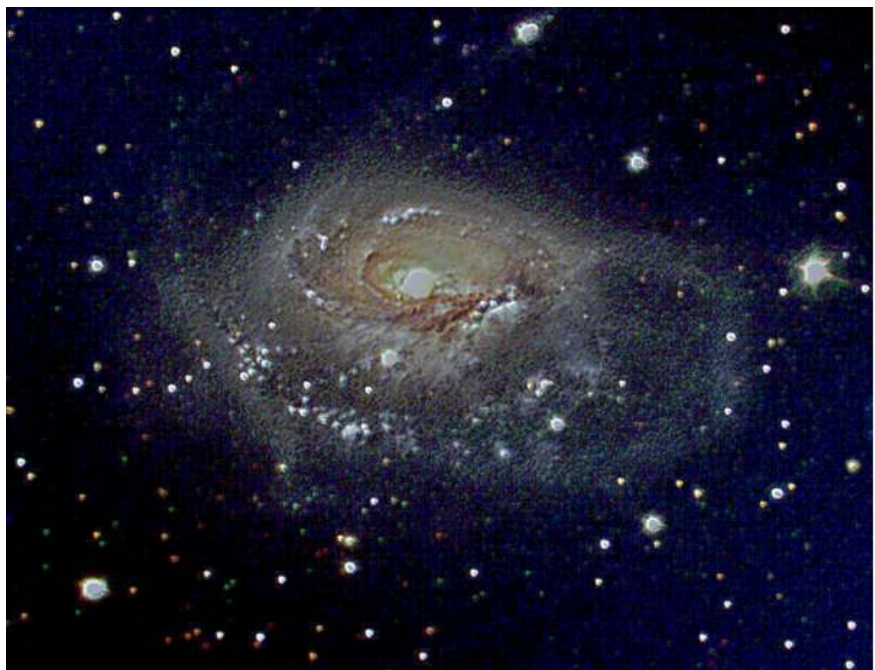

**Figure 4.** *AstroFracTool* image, with a subsequent GIMP adjustment, obtained from an image of NGC1961 galaxy (Nordic Optical Telescope Scientific Association, Jyri Näränen and Kalle Torstensson).

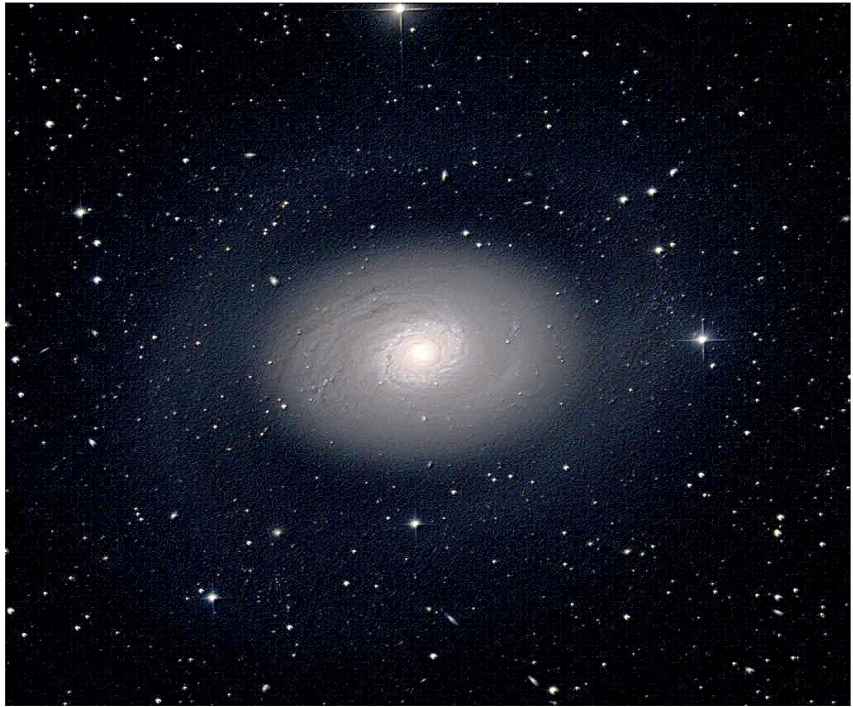

**Figure 5.** A*stroFracTool* image, with a subsequent GIMP adjustment, obtained from an image of M94 galaxy, by Hillary Mathis, N.A.Sharp/NOAO/AURA/NSF.

9